\documentclass[aps,prd,showpacs,floatfix,nofootinbib,superscriptaddress,twocolumn,10pt]{revtex4}
\usepackage{graphicx}
\usepackage{bm}
\usepackage{subfigure}
\usepackage{amssymb}
\usepackage{color,soul}
\usepackage{graphicx}
\usepackage{multirow}
\usepackage{amsmath}
\graphicspath{{figure/}}
\DeclareGraphicsExtensions{.pdf,.png,.jpg}
\setulcolor{red}

\newcommand{\beq}{\begin{equation}}
\newcommand{\eeq}{\end{equation}}
\newcommand{\beqa}{\begin{eqnarray}}
\newcommand{\eeqa}{\end{eqnarray}}

\begin{document}

\title{The Possibility of Mirror Planet as Planet Nine in Solar System}

\author{Pei Wang}
\affiliation{Key Laboratory of Dark Matter and Space Astronomy, Purple Mountain Observatory, Chinese Academy of Sciences, Nanjing 210023, China}
\affiliation{School of Astronomy and Space Science, University of Science and Technology of China, Hefei, Anhui 230026, China}

\author{Yu-Chen Tang}
\affiliation{Key Laboratory of Dark Matter and Space Astronomy, Purple Mountain Observatory, Chinese Academy of Sciences, Nanjing 210023, China}
\affiliation{School of Astronomy and Space Science, University of Science and Technology of China, Hefei, Anhui 230026, China}

\author{Lei Zu \footnote{Corresponding author: zulei@pmo.ac.cn}}
\affiliation{Key Laboratory of Dark Matter and Space Astronomy, Purple Mountain Observatory, Chinese Academy of Sciences, Nanjing 210023, China}
\affiliation{School of Astronomy and Space Science, University of Science and Technology of China, Hefei, Anhui 230026, China}

\author{Yuan-Yuan Chen}
\affiliation{Purple Mountain Observatory, Chinese Academy of Sciences, Nanjing 210023, China}
\affiliation{Key Laboratory of Planetary Sciences, Chinese Academy of Sciences, Nanjing 210023, China}
\affiliation{CAS Center for Excellence in Comparative Planetology, Hefei, Anhui 230026, China}

\author{Lei Feng \footnote{Corresponding author: fenglei@pmo.ac.cn}} 
\affiliation{Key Laboratory of Dark Matter and Space Astronomy, Purple Mountain Observatory, Chinese Academy of Sciences, Nanjing 210023, China}
\affiliation{Joint Center for Particle, Nuclear Physics and Cosmology,  Nanjing University -- Purple Mountain Observatory,  Nanjing  210093, China}


\begin{abstract}

A series of dynamical anomalies in the orbits of distant trans-Neptunian objects points to a new celestial body (usually named Planet Nine) in the solar system. In this draft, we point out that a mirror planet captured from the outer solar system or formed in the solar system is also a possible candidate. The introduction of the mirror matter model is due to an unbroken parity symmetry and is a potential explanation for dark matter. This mirror planet has null or fainter electromagnetic counterparts with a smaller optical radius and might be explored through gravitational effects.

\end{abstract}

\pacs{}
\maketitle

\section{introduction}

Over the past two decades, many observations have unveiled an abnormal structure of distant trans-Neptunian objects (TNOs), which cannot be explained by the known eight-planet solar system alone \cite{brown2004,trujillo2014}. While a new ninth planet in our Solar System could account for this dynamical phenomenon \cite{batygin2016,batygin2019}. Planet Nine (P9) is predicted with a mass of $\sim$5--10 $M_\oplus$, semi-major axis of $a_9$$\sim$400--800 AU and eccentricity between $e_9$$\sim$0.2--0.5 \cite{batygin2019}. The probability of P9 has been discussed for scattering the planet from a smaller orbit, ejecting it from the solar system or capturing it from another system as a free-floating planet \cite{batygin2019,li2016,mustill2016}. It would indicate that the current theories for forming planets need to be updated. Further studies have also been carried out for the possible magnitudes \cite{linder2016}, the mass and radius \cite{schneider2017} of the P9.

Considering the absence of an apparent optical counterpart by now, the authors in Ref.~\cite{scholtz2020} discussed the possibility of a primordial black hole (BH) as the P9.
It aims to explain two anomalies: the orbits of TNOs and the microlensing events in the Optical Gravitational Lensing Experiment (OGLE) with an object mass of M$ \sim$0.5--20 $M_\oplus$ \cite{mroz2017, niikura2019}.
However, the primordial BH is not the only dark component in the mass range of approximately $M_\oplus$ beyond the standard models 
\cite{2022arXiv220809566K,tolos2015,foot2014,foot1999,foot2001}.
{{Other hybrid objects were also discussed by several groups}. For example, in Ref. \cite{leung2011dark,perez2022cooling}, the authors proposed the possibility of small neutron stars which consist of a small compact matter core in a dark matter halo. However, considering the minimum mass of neutron stars $\sim$0.1~$M_\odot$ \cite{haensel2002equation} is higher than the mass of P9, it is out of consideration here.}   

Here, we discuss the possibility of a mirror planet acting as the P9 which is formed by mirror dark matter, another possible dark matter candidate. In this scenario, the mirror sector is the exact copy of the standard model sector to keep the full Poincar$\rm \acute{e}$ symmetry \cite{foot2014}. The non-gravitational interactions between the two sectors are negligible, such that the mirror part could constitute the non-baryonic dark matter in the universe. However, the mirror particles hold the same interactions as the ordinary parts, which predict a similar structure such as dark stars, dark neutron stars, and even dark planets \cite{curtin2020,sandin2009,foot2002,2022JHEP...07..059H}. Mirror planets in the solar system have also been discussed in \cite{foot2001mirror,foot2003}. 
{{In Ref.}
~\cite{foot1999}, the authors argued that  several close-orbiting large mass exoplanets may be made of mirror particles, i.e., they are mirror planets. Another fascinating hypothesis is that a mirror matter
space body may be responsible for the Tunguska explosion~\cite{foot2001mirror1}.}
From our calculation, we found that 
there are enough mirror particles captured in the solar system to form a dark planet in this mass range. In addition, this dark planet may come from the gravitational capture of the solar system, and the capturing rate is compatible with the ordinary free-floating planet. This dark planet also reflects sunlight as it accretes ordinary matter from the environment~\cite{curtin2020}, and its magnitude was also calculated in this draft.

This paper is organized as follows: In Section \ref{sec2}, we introduce the mirror dark matter model and discuss the probability of the mirror planet in our solar system. Then, in \mbox{Section \ref{sec3}}, we discuss the possible distinguishable observation features for the mirror planet. The conclusions are summarized in the final section.

\section{mirror planet in solar system}
\label{sec2}

Each known particle has a mirror partner in the mirror matter model. This hidden sector includes the new $Z_2$ symmetry, thus, explaining 
the left-right hand symmetry that is broken in the standard model \cite{Barbieri:2005ri,Chacko:2005un,Chacko:2005vw,foot2014}. 
If this $Z_2$ symmetry is entirely unbroken, the hidden sector will interact amongst themselves, leading to a similar dynamic with the ordinary particles. {{These mirror particles,} although put forward for the symmetry broken problem initially, are also good candidates for dark matter in astrophysics because of the weak coupling with ordinary particles. In addition, mirror particles have a strong interaction within the mirror world, which leads to the mirror-particle-abundant phenomenon 
	as a kind of self-interacting dark matter \cite{foot2014}.} Celestial bodies in the normal world, such as halos, stars and planets, are also possible in the mirror world \cite{foot2014,curtin2020,foot2001mirror,2022JHEP...05..050B}.

The only allowed interaction between mirror particles and the normal part is through the kinetic mixing or Higgs mixing \cite{foot2014}:
\begin{equation}
L_{mix}=\frac{\epsilon}{2} F^{\mu \nu}F^{\prime}_{\mu \nu} + \lambda \phi \phi^{\dagger} \phi^{\prime} \phi^{\prime \dagger},
\label{lmix}
\end{equation}
where $F^{\mu\nu}(F_{\mu\nu}^\prime)$  describes the ordinary (mirror) photon tensor and $\phi$($\phi^{\prime}$) describes the ordinary (mirror) Higgs. In this work, we only consider the kinetic mixing term, which has already been related to many astronomical and cosmological observations \cite{foot2014,2021JHEP...11..198C,zu2021}. {{With this mixing,} where $\epsilon$ is a constant of approximately 
	$\sim$10$^{-10}$ to explain the small scale structure \cite{foot2014}, the solar system could capture enough mirror particles to form the P9 in history.} Here, we adopt the Equation~(196) in Ref.~\cite{foot2014} to calculate the mass of the mirror planet, which could have been accumulated in the history of the solar system formation:

\begin{equation}
M^{\prime} \sim 10 M_\oplus \frac{\pi R_0^2}{(0.01\ \mathrm{pc})^2}\frac{T}{10^5\ \mathrm{yr}} ,
\label{m_captured}
\end{equation}
where $M_\oplus$ is the mass of Earth, $R_0$ is the effective captured radius for the nebular disk and $T$ is the formation period of the protosun and nebular disk. Thus, in the formation period, even with no mirror components at first, our solar system can accumulate a substantial amount of mirror particles, which might be enough for the P9. 

Furthermore, the P9 might also come from the outer solar system, such as the capture of a free-floating planet \cite{batygin2019,li2016}. The capture rate can be expressed as \cite{scholtz2020}:
\begin{equation}
\Gamma = \int n_0 F(v+v_{\odot,r})\frac{d\sigma}{dv}v dv ,
\label{captured_rate}
\end{equation}
where $F(v)$ and $n_0$ are the velocity distribution and the number density of the objects to be captured, $d\sigma/dv$ is the differential capture cross-section, and $v_{\odot,r}$ is the velocity of the Sun concerning the rest frame {{\bf r}} 
 of the planets to be captured.

Assuming that the mirror planets hold the same velocity distribution as an ordinary free-floating planet since they share the same dynamical process, the capture rate ratio between the two parts is proportional to the local density\cite{scholtz2020}:

\begin{equation}
\frac{\Gamma_{FFP}}{\Gamma_{MP}} \sim \frac{1}{5} f\frac{n_{FFP}}{n_{MP}} ,
\label{captured_ratio}
\end{equation}
where $n_{FFP}$  ($n_{MP}$) is the local density of free-floating ordinary planets (mirror planets), and $f$ is the fraction of mirror particles as compact objects such as mirror stars and mirror planets, other than the plasma particles. The constant $1/5$ is the density ratio between baryon and dark matter ($\Omega_{b}/\Omega_{DM}$). In Ref.~\cite{beradze2019}, the authors expected that the mirror world has almost the same structure as the ordinary---meaning $f$$\sim$1---to explain the gravitational wave signals without electromagnetic counterparts. 
As shown in Ref.~\cite{foot2014}, the scenario was found that $10\%$$\sim$$20\%$ of the Milky Way halo are mirror stars consistent with the Massive Compact Halo Objects observation. Thus, we can expect the capture rate of the mirror planet with Equation (\ref{captured_ratio}), if it exists, to be at least comparable with the ordinary free-floating planet.

\section{observation of mirror P9}
\label{sec3}

The magnitudes, mass and radius of the P9 have been discussed in \cite{linder2016,schneider2017}. Future telescopes, such as the Large Synoptic Survey Telescope~\cite{LSST} and Thirty Meter Telescope (TMT) ~\cite{TMT1,TMT2}, are expected to search this P9  \cite{linder2016}. The detection of stellar occultations and gravitational effects can also be used to determine the mass and radius of P9~\cite{schneider2017}. However, if the P9 is completely a mirror planet, only gravitation effects, such as microlensing and time delay, are expected to be observed. The optical observation for the direct detection or the stellar occultations would have a null result, which predicts a unique observation feature compared to the ordinary planet. However, that is only correct in the case where the interaction between two worlds is weak enough.

Even if we consider the accretion history of the mirror planet, the amount of the accreted ordinary particles should also be low if the initial is a pure mirror planet in these orbits~\cite{foot2014}. 
When we assume that the P9 is in the hydrostatic equilibrium, the balance between pressure gradient and gravity can be described as
\cite{foot2001}:
\begin{equation}
    \frac{d P}{d r}=-\rho^{(o)}g ,
    \label{eq:balance}
\end{equation}
where $\rho^{(o)}$ is the density of the ordinary matter in the P9. In addition, $g$ is estimated as:
\begin{equation}
    g\approx \frac{4\pi G\rho^{(m)}r}{3} ,
    \label{eq:g}
\end{equation}
where $\rho^{(m)}$ is the density of the mirror matter in the P9. Following \cite{foot2001}, we assume that the ordinary matter is mainly molecular hydrogen ($H_2$), which is natural for the local environment. In addition, the ordinary matter gas should obey the ideal gas law:
\begin{equation}
    P= \frac{\rho^{(o)}k T}{m_{H_2}} 
    \label{eq:gas_law} ,
\end{equation}
where k is Boltzmann's constant and $m_{H_2}$ is the molecular hydrogen mass. Substituting Equations~\eqref{eq:g} and \eqref{eq:gas_law} into Equation~\eqref{eq:balance} and solving the resulting differential equation, we obtain the solution \cite{foot2001}:
\begin{equation}
    \frac{\rho^{(o)}(r)}{\rho^{(o)}(0)}=\frac{T(0)}{T(r)} e^{-r^{2} / R_{x}^{2}} \ \text { for } r<R_{m} ,
    \label{eq:solution}
\end{equation}
where
\begin{equation}
    R_x\equiv \sqrt{\frac{3k}{4\pi m_p G \rho^{(m)}\lambda} } ,
\end{equation}
and 
\begin{equation}
    \lambda \equiv \frac{1}{r^2}\int_{0}^{r} \frac{1}{T(r^{'})}dr^{'2},
\end{equation}
where $R_m$ is the mirror matter radius. If we consider that the temperature is the same at different radii of the P9, then Equation~\eqref{eq:solution} will become:
\begin{equation}
    \rho^{(o)}(r)= \rho^{(o)}(0)e^{-r^{2} / R_{x}^{2}} \ \text { for } r<R_{m}.
    \label{eq:density_r}
\end{equation}

The temperature of the mirror planet will increase as the radius  decreases, which means that $\lambda < 1/T_s$. 
 Then, the lower limit for $R_x$ can be obtained:
\begin{equation}
    R_{x} \geq 5 \times 10^{8} \sqrt{\frac{T_{s}} {10^{3}\ \rm{K} }\frac{1\ \mathrm{g} / \mathrm{cm}^{3} }{\rho^{(m)}}}\  \mathrm{cm},
\end{equation}
where $T_s$ is the surface temperature of the ordinary matter. If we consider 
$\rho^{(m)}$$\sim$1~$\mathrm{g}/\mathrm{cm}^{3} $, we then estimate that
\begin{equation}
    R_{x} \sim 5 \times 10^{8} \sqrt{T_{s} / 10^{3} \ \mathrm{K}}\  \mathrm{cm}.
    \label{eq:R_x}
\end{equation}

A simple calculation can obtain the mass of the ordinary matter in the P9:
\begin{equation}
    \begin{aligned}
      M^{(o)}&=\int_{0}^{R_p} \rho^{(o)}(r) 4\pi r^{2} d r \\
             &=\int_{0}^{R_p} \rho^{(o)}(0)e^{-r^{2} / R_{x}^{2}} 4\pi r^{2} dr\\
             &=\int_{0}^{\infty } \rho^{(o)}(0)e^{-r^{2} / R_{x}^{2}} 4\pi r^{2} dr\\
             &-
             \int_{R_p}^{\infty } \rho^{(o)}(0)e^{-r^{2} / R_{x}^{2}} 4\pi r^{2} dr,
    \end{aligned}
    \label{eq:M}
\end{equation}
{{where} $R_p$ is the effective radius of the ordinary matter defined as $\rho(R_p)$$\sim$10$^{-7} \mathrm{g/cm^3}$~\cite{hubbard2001} and then the second term in Equation~\eqref{eq:M}~can be ignored. Thus, the mass of the ordinary matter can be estimated as follows:}
\begin{equation}
    \begin{aligned}
          M^{(o)}&\approx \int_{0}^{\infty } \rho^{(o)}(0)e^{-r^{2} / R_{x}^{2}} 4\pi r^{2} dr\\
            &\approx 4\pi\rho^{(o)}(0) \int_{0}^{\infty} e^{-r^{2} / R_{x}^{2}} r^{2} dr\\
             &\approx \pi^{\frac{3}{2} }R_x^3\rho^{(o)}(0).
    \end{aligned}
\end{equation}
{In addition,} 
 the effective optical radius $R_p$ can be derived from the equation: 
\begin{equation}
    \begin{aligned}
    \rho^{(o)}(R_p)&= \rho^{(o)}(0)e^{-R_p^{2} / R_{x}^{2}}\\
    &\approx \frac{M^{(o)}}{\pi^{3/2}R_x^3} e^{-R_p^{2} / R_{x}^{2}}\\
    &\sim 10^{-7}\ \mathrm{g/cm^3}.
    \end{aligned}
    \label{eq:density_critical}
\end{equation}

Thus, the optical effects of the P9 come from the reflection of sunlight on the planet's surface. The luminosity of the P9 can be described as:
\begin{equation}
    \begin{aligned}
    L_{max}&=\alpha \frac{L_{\odot}}{4\pi r_p^2}\times  \pi R_p^2\\
           &=\alpha \frac{L_{\odot}}{4} \frac{R_p^2}{r_p^2},
    \end{aligned}
    \label{eq:luminosity}
\end{equation}
where $L_{\odot}$ is the sun's luminosity, $r_p$ is the planet's distance from the sun, and $\alpha$ is the albedo of the P9. Referring to the albedo of the planets in the solar system, we chose 0.3 as the value of $\alpha$. Then we obtain the apparent magnitude by the luminosity:
\begin{equation}
    m\approx M_{\odot}-2.5\lg\frac{L_{max}}{L_{\odot}} -5+5\lg r_p .
    \label{eq:magnitude}
\end{equation}

\begin{figure}
\centering
\includegraphics[scale=0.45]{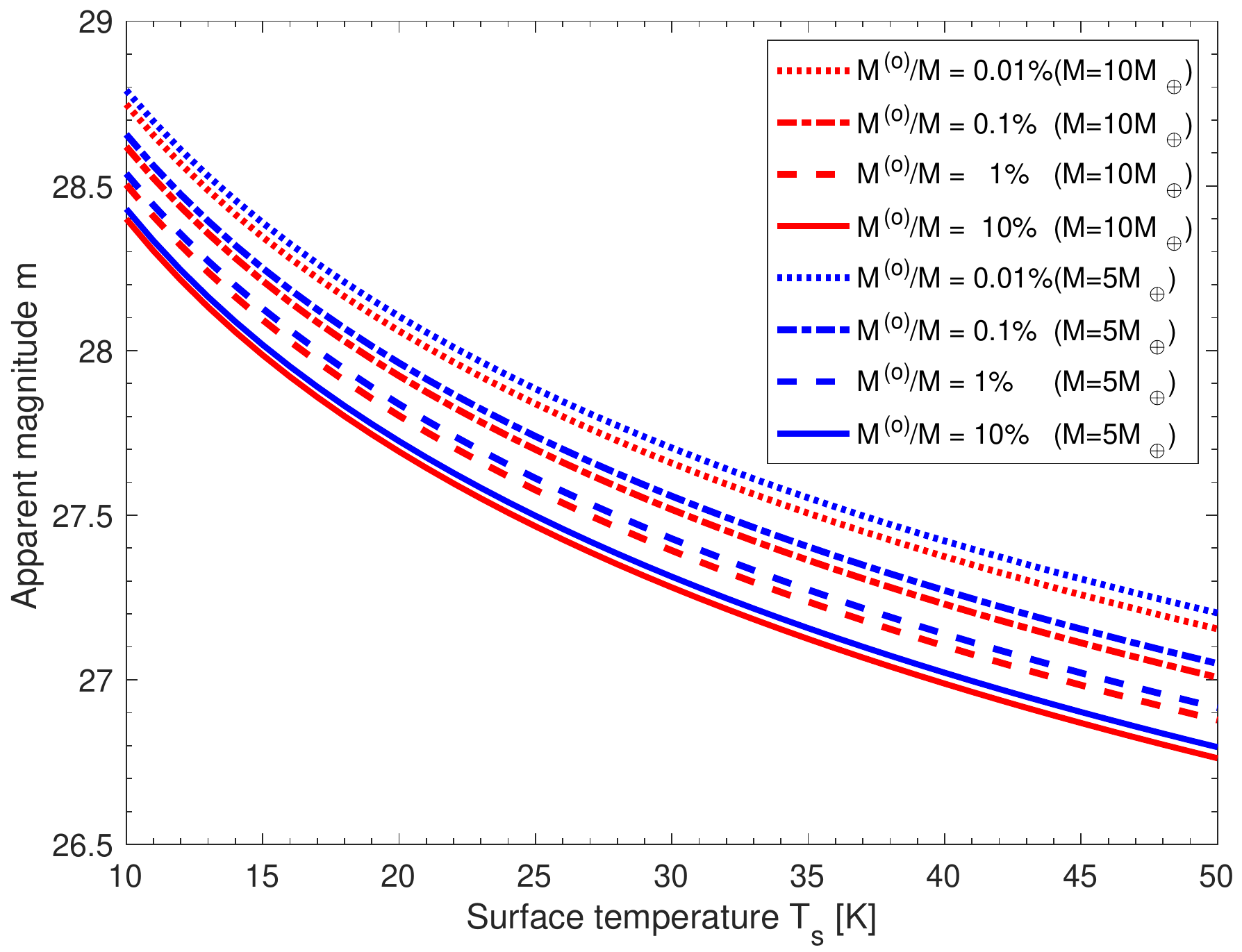}
\caption{The predicted apparent magnitude of the P9. M is the total mass of the P9.\label{fig:magnitude}}
\end{figure}

Considering that the temperature of the CMB is about $3~\mathrm{K}$ and the surface temperature of Neptune, which is the farthest known planet from the sun in the solar system, is about $55~\mathrm{K}$, the temperature of the P9 is assumed to be $10~\mathrm{K}$$\sim$50~$\mathrm{K}$.
From Equation~\eqref{eq:R_x} and Equations~\eqref{eq:density_critical}--\eqref{eq:magnitude}, we can obtain the relationship among m, $T_s$ and $M^{(o)}$. The results are shown in Figure \ref{fig:magnitude}. In the calculations, we assume that the total mass of P9 is about {{5 and 10} $M_\oplus$}, and its distance is estimated to be about 700 AU following Ref. \cite{batygin2016}.
We have shown the relationship between apparent magnitude and temperature, which depends on the mass ratio of the ordinary matter, accounting for $0.01\%$, $0.1\%$, $1\%$ and $10\%$ of the total mass. The luminosity for these situations is far beyond the detection ability of the current and near-future telescopes. Thus, we predict the null result of optical observations, and only its gravitational effect is a promising approach for detecting such a mirror planet.

{In addition,} the existence of P9 is able to be detected through gravitational microlensing in the near future. In Ref.~\cite{philippov2016}, the authors have discussed the gravitational lensing effects for P9 and concluded that microlensing is possible while strong gravitational lensing is impossible. In principle, we cannot discriminate mirror planets from ordinary planets through microlensing observations since they have the same gravitational interaction. However, occultation observation is an effective method to distinguish the case.
Following Ref.~\cite{schneider2017}, the probability of P9's occultation can be described as $P_{occ}=N_*A_9$, where $N_*$ is the number of stars per $\mathrm{arcsec^2}$ and $A_9$ is the area of the sky band swept by P9. $A_9$ is about 24 $ \mathrm{arcsec^2}$ for ordinary P9  with a radius of $\sim$$3R_\oplus$~\cite{linder2016} at 1000 AU. However, $A_9$ for the mirror planet is about $10^{-2}\  \mathrm{arcsec^2}$ which is nearly two orders less than that for the ordinary planet.
Then, its occultation probability is also much smaller correspondingly, far less than the detection ability in the near future. Two cases have the same observation effects in  microlensing but with totally different results in star occultation probability, which makes the dark component distinguishable from the ordinary planet.  

\section{conclusion}
\label{sec4}
In this draft, we explored the probability of a mirror planet as P9 and found that it could explain the anomalous orbits 
of TNOs in the solar system with a mass of $\sim$5--10 $M_\oplus$. In the formation history of our solar system, the captured mirror particles are large enough to form a planet in this mass range. In addition, the capture rate of a mirror planet is comparable with that of an ordinary free-floating planet. {{This mirror planet is} optically faint, as shown in Figure \ref{fig:magnitude}. Moreover, it has the same microlensing effects but a smaller star occultation probability compared with normal P9. If P9 is captured in the future with a much smaller star occultation and fainter counterparts, it may favor the probability of a mirror world.}

{\bf Acknowledgments}
This work is supported by the National Natural Science Foundation of China (Grants No. 11773075) and the Youth Innovation Promotion Association of Chinese Academy of Sciences (Grant No. 2016288).



\begin{thebibliography}{999}

\bibitem[Brown \em{et~al.}(2004)Brown, Trujillo, and Rabinowitz]{brown2004}
Brown, M.E.; Trujillo, C.; Rabinowitz, D.
\newblock Discovery of a candidate inner Oort cloud planetoid.
\newblock {\em  Astrophys. J.} {\bf 2004}, {\em 617},~645.

\bibitem[Trujillo and Sheppard(2014)]{trujillo2014}
Trujillo, C.A.; Sheppard, S.S.
\newblock A Sedna-like body with a perihelion of 80 astronomical units.
\newblock {\em Nature} {\bf 2014}, {\em 507},~471--474.

\bibitem[Batygin and Brown(2016)]{batygin2016}
Batygin, K.; Brown, M.E.
\newblock Evidence for a distant giant planet in the solar system.
\newblock {\em  Astron. J.} {\bf 2016}, {\em 151},~22.

\bibitem[Batygin \em{et~al.}(2019)Batygin, Adams, Brown, and
  Becker]{batygin2019}
Batygin, K.; Adams, F.C.; Brown, M.E.; Becker, J.C.
\newblock The planet nine hypothesis.
\newblock {\em Phys. Rep.} {\bf 2019}, {\em 805},~1--53.

\bibitem[Li and Adams(2016)]{li2016}
Li, G.; Adams, F.C.
\newblock Interaction cross sections and survival rates for proposed solar
  system member planet nine.
\newblock {\em  Astrophys. J. Lett.} {\bf 2016}, {\em 823},~L3.

\bibitem[Mustill \em{et~al.}(2016)Mustill, Raymond, and Davies]{mustill2016}
Mustill, A.J.; Raymond, S.N.; Davies, M.B.
\newblock Is there an exoplanet in the Solar system?
\newblock {\em Mon. Not. R. Astron. Soc. Lett.} {\bf
  2016}, {\em 460},~L109--L113.

\bibitem[Linder and Mordasini(2016)]{linder2016}
Linder, E.F.; Mordasini, C.
\newblock Evolution and magnitudes of candidate Planet Nine.
\newblock {\em Astron. Astrophys.} {\bf 2016}, {\em 589},~A134.

\bibitem[Schneider(2017)]{schneider2017}
Schneider, J.
\newblock Measuring the radius and mass of Planet Nine.
\newblock {\em Publ. Astron. Soc. Pac.} {\bf
  2017}, {\em 129},~104401.

\bibitem[Scholtz and Unwin(2020)]{scholtz2020}
Scholtz, J.; Unwin, J.
\newblock What if Planet 9 is a primordial black hole?
\newblock {\em Phys. Rev. Lett.} {\bf 2020}, {\em 125},~051103.

\bibitem[Mr{\'o}z \em{et~al.}(2017)Mr{\'o}z, Udalski, Skowron, Poleski,
  Koz{\l}owski, Szyma{\'n}ski, Soszy{\'n}ski, Wyrzykowski, Pietrukowicz,
  Ulaczyk, et~al.]{mroz2017}
Mr{\'o}z, P.; Udalski, A.; Skowron, J.; Poleski, R.; Koz{\l}owski, S.;
  Szyma{\'n}ski, M.K.; Soszy{\'n}ski, I.; Wyrzykowski, {\L}.; Pietrukowicz, P.;
  Ulaczyk, K.;  et~al.
\newblock No large population of unbound or wide-orbit Jupiter-mass planets.
\newblock {\em Nature} {\bf 2017}, {\em 548},~183--186.

\bibitem[Niikura \em{et~al.}(2019)Niikura, Takada, Yokoyama, Sumi, and
  Masaki]{niikura2019}
Niikura, H.; Takada, M.; Yokoyama, S.; Sumi, T.; Masaki, S.
\newblock Constraints on Earth-mass primordial black holes from OGLE 5-year
  microlensing events.
\newblock {\em Phys. Rev. D} {\bf 2019}, {\em 99},~083503.

\bibitem[{Kitabayashi}(2022)]{2022arXiv220809566K}
{Kitabayashi}, T.
\newblock {Primordial black holes and mirror dark matter}.
\newblock {\em arXiv} {\bf 2022}, arXiv:2208.09566.

\bibitem[Tolos and Schaffner-Bielich(2015)]{tolos2015}
Tolos, L.; Schaffner-Bielich, J.
\newblock Dark compact planets.
\newblock {\em Phys. Rev. D} {\bf 2015}, {\em 92},~123002.

\bibitem[Foot(2014)]{foot2014}
Foot, R.
\newblock Mirror dark matter: Cosmology, galaxy structure and direct detection.
\newblock {\em Int. J. Mod. Phys. A} {\bf 2014}, {\em
  29},~1430013.

\bibitem[Foot(1999)]{foot1999}
Foot, R.
\newblock Have mirror planets been observed?
\newblock {\em Phys. Lett. B} {\bf 1999}, {\em 471},~191--194.

\bibitem[Foot(2001)]{foot2001}
Foot, R.
\newblock Are mirror worlds opaque?
\newblock {\em Phys. Lett. B} {\bf 2001}, {\em 505},~1--5.

\bibitem[Leung \em{et~al.}(2011)Leung, Chu, and Lin]{leung2011dark}
Leung, S.C.; Chu, M.C.; Lin, L.M.
\newblock Dark-matter admixed neutron stars.
\newblock {\em Phys. Rev. D} {\bf 2011}, {\em 84},~107301.

\bibitem[P{\'e}rez-Garc{\'\i}a \em{et~al.}(2022)P{\'e}rez-Garc{\'\i}a,
  Grigorian, Albertus, Barba, and Silk]{perez2022cooling}
P{\'e}rez-Garc{\'\i}a, M.{\'A}.; Grigorian, H.; Albertus, C.; Barba, D.; Silk,
  J.
\newblock Cooling of Neutron Stars admixed with light dark matter: A case
  study.
\newblock {\em Phys. Lett. B} {\bf 2022}, {\em 827},~136937.

\bibitem[Haensel \em{et~al.}(2002)Haensel, Zdunik, and
  Douchin]{haensel2002equation}
Haensel, P.; Zdunik, J.; Douchin, F.
\newblock Equation of state of dense matter and the minimum mass of cold
  neutron stars.
\newblock {\em Astron. Astrophys.} {\bf 2002}, {\em 385},~301--307.

\bibitem[Curtin and Setford(2020)]{curtin2020}
Curtin, D.; Setford, J.
\newblock Signatures of mirror stars.
\newblock {\em J. High Energy Phys.} {\bf 2020}, {\em 2020},~1--44.

\bibitem[Sandin and Ciarcelluti(2009)]{sandin2009}
Sandin, F.; Ciarcelluti, P.
\newblock Effects of mirror dark matter on neutron stars.
\newblock {\em Astropart. Phys.} {\bf 2009}, {\em 32},~278--284.

\bibitem[Foot \em{et~al.}(2002)Foot, Ignatiev, and Volkas]{foot2002}
Foot, R.; Ignatiev, A.Y.; Volkas, R.
\newblock Do “isolated” planetary mass objects orbit mirror stars?
\newblock {\em Astropart. Phys.} {\bf 2002}, {\em 17},~195--198.

\bibitem[{Howe} \em{et~al.}(2022){Howe}, {Setford}, {Curtin}, and
  {Matzner}]{2022JHEP...07..059H}
{Howe}, A.; {Setford}, J.; {Curtin}, D.; {Matzner}, C.D.
\newblock {How to search for mirror stars with Gaia}.
\newblock {\em J. High Energy Phys.} {\bf 2022}, {\em 2022},~59.
\newblock {\url{https://doi.org/10.1007/JHEP07(2022)059}}.

\bibitem[Foot and Silagadze(2001)]{foot2001mirror}
Foot, R.; Silagadze, Z.
\newblock Do mirror planets exist in our solar system?
\newblock {\em arXiv} {\bf 2001}, \emph{preprint}. arXiv:astro-ph/0104251.

\bibitem[Foot and Mitra(2003)]{foot2003}
Foot, R.; Mitra, S.
\newblock Mirror matter in the solar system: New evidence for mirror matter
  from Eros.
\newblock {\em Astropart. Phys.} {\bf 2003}, {\em 19},~739--753.

\bibitem[Foot(2001)]{foot2001mirror1}
Foot, R.
\newblock The mirror world interpretation of the 1908 Tunguska event and other
  more recent events.
\newblock {\em arXiv} {\bf 2001}, \emph{preprint}. arXiv:hep-ph/0107132.

\bibitem[Barbieri \em{et~al.}(2005)Barbieri, Gregoire, and
  Hall]{Barbieri:2005ri}
Barbieri, R.; Gregoire, T.; Hall, L.J.
\newblock {Mirror World at the Large Hadron Collider} {2005}.
\newblock Available online: \url{http://xxx.lanl.gov/abs/hep-ph/0509242} {(accessed on 1 September 2022).} 


\bibitem[Chacko \em{et~al.}(2006{\natexlab{a}})Chacko, Goh, and
  Harnik]{Chacko:2005un}
Chacko, Z.; Goh, H.S.; Harnik, R.
\newblock {A Twin Higgs model from left-right symmetry}.
\newblock {\em JHEP} {\bf 2006}, {\em 01},~108.
\newblock {\url{https://doi.org/10.1088/1126-6708/2006/01/108}}.

\bibitem[Chacko \em{et~al.}(2006{\natexlab{b}})Chacko, Nomura, Papucci, and
  Perez]{Chacko:2005vw}
Chacko, Z.; Nomura, Y.; Papucci, M.; Perez, G.
\newblock {Natural little hierarchy from a partially goldstone twin Higgs}.
\newblock {\em JHEP} {\bf 2006}, {\em 01},~126. 
\newblock {\url{https://doi.org/10.1088/1126-6708/2006/01/126}}.

\bibitem[{Bansal} \em{et~al.}(2022){Bansal}, {Kim}, {Kolda}, {Low}, and
  {Tsai}]{2022JHEP...05..050B}
{Bansal}, S.; {Kim}, J.H.; {Kolda}, C.; {Low}, M.; {Tsai}, Y.
\newblock {Mirror twin Higgs cosmology: Constraints and a possible resolution
  to the H$_{0}$ and S$_{8}$ tensions}.
\newblock {\em J. High Energy Phys.} {\bf 2022}, {\em 2022},~50. 
\newblock {\url{https://doi.org/10.1007/JHEP05(2022)050}}.

\bibitem[{Chacko} \em{et~al.}(2021){Chacko}, {Curtin}, {Geller}, and
  {Tsai}]{2021JHEP...11..198C}
{Chacko}, Z.; {Curtin}, D.; {Geller}, M.; {Tsai}, Y.
\newblock {Direct detection of mirror matter in Twin Higgs models}.
\newblock {\em J. High Energy Phys.} {\bf 2021}, {\em 2021},~198. 
\newblock {\url{https://doi.org/10.1007/JHEP11(2021)198}}.

\bibitem[Zu \em{et~al.}(2021)Zu, Yuan, Feng, and Fan]{zu2021}
Zu, L.; Yuan, G.W.; Feng, L.; Fan, Y.Z.
\newblock Mirror dark matter and electronic recoil events in XENON1T.
\newblock {\em Nucl. Phys. B} {\bf 2021}, {\em 965},~115369.

\bibitem[Beradze and Gogberashvili(2019)]{beradze2019}
Beradze, R.; Gogberashvili, M.
\newblock LIGO signals from the mirror world.
\newblock {\em Mon. Not. R. Astron. Soc.} {\bf 2019},
  {\em 487},~650--652.

\bibitem[LSS()]{LSST}
{Available online: \url{http://www.lsst.org} {(accessed on 1 September 2022).}}

\bibitem[TMT()]{TMT1}
Available online: \url{https://www.tmt.org} {(accessed on 1 September 2022).}

\bibitem[{Sanders}(2013)]{TMT2}
{Sanders}, G.H.
\newblock {The Thirty Meter Telescope (TMT): An International Observatory}.
\newblock {\em J. Astrophys. Astron.} {\bf 2013}, {\em
  34},~81--86.
\newblock {\url{https://doi.org/10.1007/s12036-013-9169-5}}.

\bibitem[Hubbard \em{et~al.}(2001)Hubbard, Fortney, Lunine, Burrows, Sudarsky,
  and Pinto]{hubbard2001}
Hubbard, W.; Fortney, J.; Lunine, J.; Burrows, A.; Sudarsky, D.; Pinto, P.
\newblock Theory of extrasolar giant planet transits.
\newblock {\em  Astrophys. J.} {\bf 2001}, {\em 560},~413.

\bibitem[Philippov and Chobanu(2016)]{philippov2016}
Philippov, J.; Chobanu, M.
\newblock Nemesis, Tyche, Planet Nine Hypotheses. I. Can We Detect the Bodies
  Using Gravitational Lensing?
\newblock {\em Publ. Astron. Soc. Aust.} {\bf
  2016}, {\em {33}}. 


\end{thebibliography}
\end{document}